\documentclass[twocolumn,showpacs,preprintnumbers,amsmath,amssymb]{revtex4}
\usepackage{graphicx}
\usepackage{dcolumn}
\usepackage{bm}

\begin{document}

\preprint{APS/123-QED}

\title{Threshold quantum cryptograph based on Grover's algorithm}

\author{Jian-Zhong Du$^{1,2}$, Su-Juan Qin$^{1}$, Qiao-Yan Wen$^{1}$, and Fu-Chen Zhu$^{3}$\\
        $^1$School of Science, Beijing University of Posts and Telecommunications, Beijing, 100876, China \\
        $^2$State Key Laboratory of Integrated Services Network, Xidian University, Xi'an, 710071, China \\
        $^3$National Laboratory for Modern Communications, P.O.Box 810, Chengdu, 610041, China \\ Email: ddddjjjjzzzz@tom.com}

\date{\today}

\begin{abstract}
Grover's operator in the two-qubit case can transform a basis into
its conjugated basis. A permutation operator can transform a state
in the two conjugated bases into its orthogonal state. These
properties are included in a threshold quantum protocol.  The
proposed threshold quantum protocol  is secure based the proof that
the legitimate participators can only eavesdrop 2 bits of 3 bits
operation information on one two-qubit with error probability $3/8$.
We propose a scheme to detect the Trojan horse attack without
destroying the legal qubit.
\end{abstract}


\maketitle

\section{INTRODUCTION}
In a secure multi-party computation \cite {yao,ccd}, $n$
participants, $P_1, P_2, ... P_n ,$ compute and reveal the result of
the multi-variable function $f(x_1,x_2,...x_n)$, where $x_i$ is a
secret input provided by $P_i$. It is also necessary to preserve the
maximum privacy of each input $x_i$. The menace of input leakage
comes from eavesdroppers and the dishonest participants. In contrast
to the eavesdroppers outside, the dishonest participants have many
advantages to attack  another's input. As pointed out in Ref.\cite
{kmn}, if multi-party scheme is secure for the dishonest
participants, it is secure for any eavesdropper.

Based on the operators $I=\left(%
\begin{array}{cc}
  1 & 0 \\
  0 & 1 \\
\end{array}%
\right),U=\left(%
\begin{array}{cc}
  0 & 1 \\
  -1 & 0 \\
\end{array}%
\right)$, $H=\frac {1} {\sqrt {2}}
\left(%
\begin{array}{cc}
  1 & 1 \\
  1 & -1 \\
\end{array}%
\right)$ and $\overline{H}=\frac {1} {\sqrt {2}}
\left(%
\begin{array}{cc}
  -1 & 1 \\
  1 & 1 \\
\end{array}%
\right)$, quantum secure direct communication (QSDC) protocols \cite
{cl,dl,lm}, multiparty quantum secret sharing (MQSS) protocols \cite
{zlm,dhhz,qgwz} and threshold quantum protocol \cite {toi} have been
proposed. Lucamarini and Mancini \cite {lm} showed that the QSDC
protocols \cite {cl,dl,lm} are quasisecure to eavesdropper.

Deng et al.\cite {dhhz} showed a Trojan horse attack scheme against
MQSS protocol proposed by Ref.\cite {zlm}. A dishonest participant
prepares a multi-photon instead of a legal single-photon and sends
it to another participant. Then he measures the photons with some
photon number splitter (PNS) and detectors.  The attack  introduces
no error into the communication.

Qin et al.\cite {qgwz} showed another attack scheme against MQSS
protocol proposed by Ref.\cite {zlm}. A dishonest participant
prepares the fake state $(|01\rangle_{12}-|10\rangle_{12})/\sqrt
{2}$ and then sends the first qubit to another participant. After
receiving the first qubit operated, the dishonest participant can
know another participant's operation $I,U, H$ or $\overline{H}$ by
measuring qubits $1,2$ in the basis $\{(|01\rangle-|10\rangle)/\sqrt
{2}, (|00\rangle+|11\rangle)/\sqrt
{2},(|00\rangle-|01\rangle-|10\rangle-|11\rangle)/2,
(|00\rangle+|01\rangle+|10\rangle-|11\rangle)/2\}$. The attack also
introduces no error into the communication.

Participants can pick out a subset of the photons as the sample for
eavesdropping check. Deng et al.\cite {dhhz} proposed that
participants split each signal of the sample with a PNS and measure
the two
 signals. Qin et al.\cite {qgwz} proposed that participants replace
 the sample photons with
decoy photons.

 In a threshold quantum cryptography, assumption that
all the participants are honest is infeasible. The t-out-of-n
quantum cash threshold protocol proposed by Tokunaga et al.\cite
{toi} is not secure. With the help of the attack schemes proposed in
Ref.\cite {qgwz,dhhz}, the first participant $P_1$, called a center
in Ref.\cite {toi}, can completely eavesdrop $t-1$ secret inputs
kept by $t-1$ other participants in a issuing phase one by one, and
then reconstructs the copies of quantum cash that can pass the
checking phase.

The sample photons schemes \cite {dhhz,qgwz} can improve the
security of the threshold quantum protocol \cite {toi}. However the
number of qubits of the generated quantum state by the threshold
protocol must exceed that of the quantum state generated by the
original (nonthreshold) protocol. We do not follow this line of
argument. Instead we modify the protocol in the two-qubit quantum
operation.

A quantum computation consists of three constituents: generating
quantum states, performing unitary operations and measuring quantum
states. The dishonest participants whose number is less than
threshold number can generate the fake quantum states or perform the
fake unitary operations to attack one other's input, but before
measurement, they must reconstruct the legal quantum state to avoid
the detection. In this paper, a honest measurer is assumed .

In this paper, instead of one-qubit operators, we show that
two-qubit operators  based on Grover¡¯s algorithm \cite {long, gr}
can adapt to threshold quantum cryptography protocol. Each
participant
 does one of eight kinds
of operations  on every two-qubit as input. The dishonest
participants can eavesdrop 2 bits of 3 bits operation information on
one two-qubit at most with whether fake signal or legal signal. The
dishonest participants have to introduce $\frac {3} {8}$ error
probability into one two-qubit when they  eavesdrop maximum
information quantity 2 bits. These properties guarantee the proposed
threshold quantum protocol against an attack with a fake signal.
Moreover, since even the three-qubit Grover's algorithm has been
experimentally realized \cite {vss}, threshold quantum cryptography
protocol based on Grover's algorithm becomes highly practical for
experimental realization.

In this paper, we propose a detection scheme to distinguish  one
single-qubit from one multi-qubit  without destroying the legal
qubit. The scheme can detect one multi-qubit instead of one
single-qubit with probability $\frac {1} {2}$. So  a Trojan horse
attack \cite {dhhz,grtz} can be resisted.

 This paper is
organized as follows. Section II introduces two-qubit operators
based on Grover¡¯s algorithm \cite {gr}.  Section III proposes an
t-out-of-n quantum cash threshold protocol followed the line
sketched in \cite {toi} but with some relevant differences. Section
IV then shows security of the threshold protocol.  Section V proofs
that Trojan horse attack can be detected. Section VI then draws some
conclusions.

\section{TWO-QUBIT OPERATIONS BASED ON GROVER'S ALGORITHM}
Grover's operator \cite {gr} in the two-qubit case
\begin{eqnarray}
V=\frac {1} {2}
\left(%
\begin{array}{cccc}
  -1 & 1 & 1 & 1\\
  1 & -1 & 1 & 1\\
  1 & 1 & -1 & 1\\
  1 & 1 & 1 & -1\\
\end{array}%
\right)
\end{eqnarray}
can transform the basis $\{|00\rangle,|01\rangle,|10\rangle$ and
$|11\rangle\}$ to the basis
$\{|\overline{00}\rangle=1/2(-|00\rangle+|01\rangle+|10\rangle+|11\rangle),
|\overline{01}\rangle=1/2(|00\rangle-|01\rangle+|10\rangle+|11\rangle),
|\overline{10}\rangle=1/2(|00\rangle+|01\rangle-|10\rangle+|11\rangle)$
and
$|\overline{11}\rangle=1/2(|00\rangle+|01\rangle+|10\rangle-|11\rangle)\}$.

A permutation operator
\begin{eqnarray}
U=
\left(%
\begin{array}{cccc}
  0 & 0 & 0 & 1\\
  1 & 0 & 0 & 0\\
  0 & 1 & 0 & 0\\
  0 & 0 & 1 & 0\\
\end{array}%
\right)
\end{eqnarray}
has properties: $U|00\rangle=|01\rangle$, $U|01\rangle=|10\rangle$,
$U|10\rangle=|11\rangle$, $U|11\rangle=|00\rangle$,
$U|\overline{00}\rangle=|\overline{01}\rangle$,
$U|\overline{01}\rangle=|\overline{10}\rangle$,
$U|\overline{10}\rangle=|\overline{11}\rangle$, and
$U|\overline{11}\rangle=|\overline{00}\rangle$.  $V$ and $U$ are
commute operators.

In the threshold quantum proposed below, a center does one of eight
kinds unitary operation on two-qubit:
\begin{eqnarray}
U(00)V(0)=I, \nonumber\\
U(01)V(0)=U, \nonumber\\
U(10)V(0)=U \cdot U,  \nonumber\\
U(11)V(0)=U \cdot U \cdot U, \nonumber\\
U(00)V(1)=V, \nonumber\\
U(01)V(1)=U \cdot V, \nonumber\\
U(10)V(1)=U \cdot U \cdot V, \nonumber\\
U(11)V(1)=U \cdot U \cdot U \cdot V, \nonumber
\end{eqnarray}
where $I$ is identity operator.

\section{t-OUT-n THRESHOLD SCHEME}
We propose the t-out-of-n threshold version of quantum cash
protocol. There are three differences between our protocol and the
protocol proposed by Tokunaga et al.\cite {toi} mainly: one is the
assumption of dishonest participants instead of that of honest
participants, one is two-qubit operation instead of one-qubit
operation to resist the attack proposed by \cite {qgwz}, the other
is an additional detection to resist the Trojan horse attack \cite
{dhhz,grtz}. Following the line sketched in \cite {toi}, We describe
the scheme in detail.

\emph{Distribution phase}. In this phase, a dealer distributes
shared secrets to centers.

 (i) A dealer chooses an original secret
\begin{eqnarray}
K=(a_1,b_1,a_2,b_2,...,a_m,b_m)
\end{eqnarray}
for each banknote with $L_k$, where $L_k$ is a kind of serial number
(used as a label for $K$) and $a_i$, $b_i$ are uniformly chosen,
$a_i  \in \{00, 01, 10, 11\}, b_i \in \{0, 1\}$

(ii) The dealer then makes $n$ shares, $S_1,...,S_n$, of $K$ using
Shamir's secret sharing scheme \cite {s} over $\textbf{F}_{2^N}$ as
follows, where $N=3m$. The dealer chooses $x_j$'s for $j=1,...,n$
which are $n$ distinct, nonzero elements in $\textbf{F}_{2^N}$, and
the $x_j$'s are published with $L_K$. The dealer randomly chooses a
secret $(t -1)$th-degree polynomial $f(x)$ over $\textbf{F}_{2^N}$,
where $f(0)=\overline{K}$ (here, $\overline{K}$ is a polynomial
representation of $K$). Then, the dealer computes $S_j= f(x_j)$ for
j=1,...,n over $\textbf{F}_{2^N}$.

(iii) The dealer secretly sends $S_j$ with $L_K$ to center $P_j$ for
each $j=1,...,n$.

\emph{Precomputation phase}. In this phase, the centers compute the
preliminary information for the following collaborative procedure.
The preliminary information depends on which subset of centers is
chosen to collaborate. Here, for simplicity of description, we
assume that $t$ centers, $P_1,...,P_t$, collaborate to issue quantum
banknotes or check their validity. Note that the set of
collaborative centers can be different in each issuing or checking
phase.

(i) For each $j=1,...,t$, $P_j$ calculates and secretly stores the
following value (given by the Lagrange interpolation formula):
\begin{eqnarray}
K_j=S_j \prod_{1 \leq l \leq t,l\neq j}  \frac {x_l} {x_l-x_j}
\end{eqnarray}
over $\textbf{F}_{2^N}$. Let
\begin{eqnarray}
K^{[j]}=
(a_1^{[j]},b_1^{[j]},a_2^{[j]},b_2^{[j]},...,a_m^{[j]},b_m^{[j]})
\end{eqnarray}
be the binary representation of $K_j$ in $\textbf{F}_{2^N}$, where
$a_i^{[j]}\in \{00,01,10,11\}, b_i^{[j]}\in \{0,1\}$. Although each
secret value $S_j$ (and $K_j$) is kept in each center $P_j$ locally,
these values satisfy the following equations globally:
\begin{eqnarray}
\label{a} \overline{K}=\sum_{j=1}^{t}K_j
\end{eqnarray}
over $\textbf{F}_{2^N}$. In binary representation, Eq.(\ref{a}) can
be written as
\begin{eqnarray}
K=\oplus_{j=1}^{t}K^{[j]}
\end{eqnarray}
where $\oplus$ represents bitwise exclusive-OR. Note that even in
the following collaboration procedure, $K_j(K^{[j]})$ is kept secret
at $P_j$ and the original secret $\overline{K}(K^{[j]})$ is not
recovered.

\emph{Issuing phase}. In this phase, $t$ centers collaborate to
issue a banknote ($L_K, |\phi\rangle$). Here, we assume the $t$
centers are $P_1,...,P_t$. Hereafter, we will describe a sequential
protocol from $P_1$ to $P_t$, but the order is not essential, any
order is possible.

(i) $P_1$ generates a quantum state
\begin{eqnarray}
|\phi^{[1]}\rangle=|\psi_{a_1^{[1]},b_1^{[1]}}\rangle\otimes
|\psi_{a_2^{[1]},b_2^{[1]}}\rangle\otimes...\otimes|\psi_{a_m^{[1]},b_m^{[1]}}\rangle,
\end{eqnarray}
where $|\psi_{a_i^{[1]}, b_i^{[1]}}\rangle$ is defined as follow:
\begin{eqnarray}
|\phi_{00,0}\rangle=|00\rangle, |\phi_{01,0}\rangle=|01\rangle, \nonumber\\
|\phi_{10,0}\rangle=|10\rangle, |\phi_{11,0}\rangle=|11\rangle, \nonumber\\
|\phi_{00,1}\rangle=|\overline{00}\rangle, |\phi_{01,1}\rangle=|\overline{01}\rangle, \nonumber\\
|\phi_{10,1}\rangle=|\overline{10}\rangle,
|\phi_{11,1}\rangle=|\overline{11}\rangle.
\end{eqnarray}
The value of $b_i$ determines the kind of the basis. If $b_i$ is 0
then $a_i$ is encoded in the  basis \{$|00\rangle, |01\rangle,
|10\rangle, |11\rangle$\}; if $b_i$ is 1 then $a_i$ is encoded in
the  basis \{$|\overline{00}\rangle, |\overline{01}\rangle,
|\overline{10}\rangle, |\overline{11}\rangle$\}. The ($L_K,
|\phi^{[1]}\rangle$) is sent to $P_2$.

(ii) For each $j=2,\ldots,t$, when $P_j$ receives ($L_K,
|\phi^{[j-1]}\rangle$) from $P_{j-1}$, he detects the Trojan horse
attack and acts his secret input on $|\phi^{[j-1]}\rangle$.

Our detection scheme  is depicted in Fig. 1. To each qubit
$|d\rangle$ of $|\phi^{[j-1]}\rangle$, called data qubit, $P_j$
uniformly choices auxiliary qubit $|a\rangle \in \{|0\rangle,
|1\rangle\}$, acts \emph{Hadamard} gate on $|a\rangle$, performs one
CNOT gates on the auxiliary qubit and the data qubit (the former is
the controller and the latter is the target), performs the unitary
transformation
\begin{eqnarray}
T=\frac 1 {\sqrt {2}}
\left(%
\begin{array}{cccc}
  1 & 0 & 0 & 1\\
  0 & 1 & 1 & 0\\
  1 & 0 & 0 & -1\\
  0 & 1 & -1 & 0\\
\end{array}%
\right)
\end{eqnarray}
on the auxiliary qubit and the data qubit, and measures the
auxiliary qubit in basis $\{|0\rangle, |1\rangle\}$. To legal single
qubit $|d\rangle$, $T_{ad}\cdot CNOT_{ad}\cdot (H_a \otimes I_d)=I_a
\otimes I_d$, so the auxiliary qubit keeps. In Section V, we proof
that the detection scheme can detect a multi-qubit instead of a
single-qubit with probability $\frac {1} {2}$. If the auxiliary
qubit flips, a single-qubit must be replaced by a multi-qubit, so
$P_j$ rejects the banknote.

$P_j$ applies the following transformation $W^{[j]}$ to
$|\phi^{[j-1]}\rangle$:
\begin{eqnarray}
\label{b} W^{[j]}=U_1^{[j]}V_1^{[j]}\otimes
U_2^{[j]}V_2^{[j]}\otimes \cdots \otimes U_m^{[j]}V_m^{[j]},
\end{eqnarray}
where
\begin{eqnarray}
\label{c} U_i^{[j]}=U(a_i^{[j]}),\ \ V_i^{[j]}=V(b_i^{[j]}).
\end{eqnarray}
$P_j$ then obtains $|\phi^{[j]}\rangle$ by the unitary
transformation
\begin{eqnarray}
W^{[j]}: |\phi^{[j-1]}\rangle \mapsto |\phi^{[j]}\rangle,
\end{eqnarray}
and sends ($L_K, |\phi^{[j]}\rangle$) to $P_{j+1}$ ($P_{t+1}$ is the
user whom the banknote is issued to ).

\emph{Checking phase}. In this phase, $t$ centers collaborate to
check the validity of quantum banknote ($L_K, |\phi'\rangle$). Here,
we assume the $t$ centers are $P_1',\ldots,P_t'$. This set of $t$
centers can be different from the set of centers that collaborate to
issue the banknote. Each $P_j'$ has calculated $K^{[j]}{'}=
(a_1^{[j]}{'},b_1^{[j]}{'},a_2^{[j]}{'},b_2^{[j]}{'},...,a_m^{[j]}{'},b_m^{[j]}{'})$
in the precomputation phase. Let
$|\phi^{[0]}{'}\rangle=|\phi'\rangle$, and $P_0'$ be the shop.

(i) For each $j=1,\ldots,t$, when $P_j'$ receives ($L_K,
|\phi^{[j-1]}{'}\rangle$) from $P'_{j-1}$, he detects the Trojan
horse attack and applies $W^{[j]}{'}$ to $|\phi^{[j-1]}{'}\rangle$
[here, $W^{[j]}{'}$ is defined in the same manner as Eqs. (\ref
{b})-(\ref {c})]. $P_j'$ then obtains $|\varphi^{[j]}{'}\rangle$ by
the unitary transformation
\begin{eqnarray}
W^{[j]}{'}: |\phi^{[j-1]}{'}\rangle \mapsto
|\varphi^{[j]}{'}\rangle.
\end{eqnarray}
Additionally, $P_j'$ chooses a secret
\begin{eqnarray}
x^{[j]}{'}=(x_1^{[j]}{'},x_2^{[j]}{'},\ldots, x_m^{[j]}{'}),
\end{eqnarray}
where $x_i^{[j]}{'}$ is uniformly choosen from \{$0,1$\}. $P_j'$
then obtains $|\phi^{[j]}{'}\rangle$ by the unitary transformation
\begin{eqnarray}
V(x_1^{[j]}{'})\otimes \ldots \otimes V(x_m^{[j]}{'}):
\varphi^{[j]}{'}\rangle \mapsto |\phi^{[j]}{'}\rangle.
\end{eqnarray}
$P_j'$ sends ($L_K, |\phi^{[j]}{'}\rangle$) to $P'_{j+1}$
($P'_{t+1}$ is the trusted measurer).

(ii) Finally, the trusted measurer requires $P_j' (j=1,\ldots,m)$
 to send the $x^{[j]}{'}$ to him secretly, measures
$|\phi^{[j]}{'}\rangle$ in the basis
$(\oplus_{j=1}^{t}x_1^{[j]}{'},\ldots,\oplus_{j=1}^{t}x_m^{[j]}{'})$,
and gets the string
\begin{eqnarray}
(c_1,\ldots,c_m).
\end{eqnarray}
The trusted measurer then checks whether $c_i=00$ for all
$i=1,\ldots,m$. Even if just one result is not 00, the centers
reject the banknote.

Necessity of the trusted measurer: If ($L_K, |\phi'\rangle$) is an
invalid quantum banknote, a dishonest measurer can always deceive
the centers by announcing ($c_1, \ldots ,c_m$)=($00, \ldots ,00$).
So an honest measurer is necessary.  It is also necessary that the
trusted measurer receives the value $x^{[j]}{'}$ secretly, otherwise
the center $P_t'$ can always send the quantum states
$|\phi^{[t]}{'}\rangle=|00\rangle \otimes \ldots \otimes |00\rangle$
to deceive  the trusted measurer.

\setlength {\unitlength} {1cm}
\begin{picture}(8,4.2)
\put(0,3.5){auxiliary} \put(0,3.2){qubit}
\put(0.5,3.1){\line(1,0){1}} \put(1.5,2.8){\framebox(0.5,0.6){$H$}}
\put(2,3.1){\line(1,0){0.8}} \put(2.3,3.07){\circle*{0.2}}
 \put(5.6,3.1){\line(1,0){0.4}}
\put(6,2.8){\framebox(1,0.6)} \qbezier
[100](6.1,2.9)(6.4,3.5)(6.9,2.9) \put(6.4,2.8){\vector(1,2){0.28}}
\put(2.8,0.9){\framebox(2.8,2.4){$T$ transformation}}
\put(0,1.5){data} \put(0,1.2){qubit} \put(0.5,1.1){\line(1,0){2.3}}
\put(2.3,0.95){\line(0,1){2.25}} \put(2.3,1.1){\circle{0.3}}
 \put(5.6,1.1){\line(1,0){1.5}}
\put(0,0.5){FIG. 1.Detection scheme of Trojan horse attack.}
\end{picture}

\section{SECURITY PROOF}
The impossibility of Eve's eavesdropping in the threshold protocol
was shown using the quantum key distribution approach, following the
line sketched in \cite {sp}.

The eavesdropping is restricted to a dishonest participant in the
following.

A dishonest participant, called Bob, is an evil quantum physicist
able to build all devices that are allowed by the laws of quantum
mechanics. Her aim is to find out another participant's input and
then to reconstruct the quantum cash with  $t-2$ other
participants. Bob prepares a fake signal and sends it to a
participant, called Alice. Then from the fake signal operated by the
Alice, Bob tries to gain Alice's input.

Without loss of generality, we assume that Alice does one of eight
kinds of operations on every two-qubit with equal probability and
that every two-qubit operation is independent. So it is sufficient
to consider Bob's eavesdropping on one two-qubit.

Bob's fake signal can be presented as
$|\theta\rangle=|00\rangle(a|A\rangle+b|B\rangle+c|C\rangle+d|D\rangle)
+|01\rangle(e|A\rangle+f|B\rangle+g|C\rangle+h|D\rangle)
+|10\rangle(i|A\rangle+j|B\rangle+k|C\rangle+q|D\rangle)
+|11\rangle(m|A\rangle+n|B\rangle+r|C\rangle+s|D\rangle)$, where
$|A\rangle, |B\rangle, |C\rangle$, and  $|D\rangle$ are normalized
orthogonal states, and
$|a|^2+|b|^2+|c|^2+|d|^2+|e|^2+|f|^2+|g|^2+|h|^2
+|i|^2+|j|^2+|k|^2+|q|^2+|m|^2+|n|^2+|r|^2+|s|^2=1$. For simplicity,
we regard every probability amplitude as real number, but the
security proof is fitted for plural number. Bob sends the former
two-qubit to Alice and leaves the rest himself.

Alice encodes her input bit by applying one of eight kinds of
operations with equal probability. The state reads
\begin{eqnarray}
w=\frac {1} {8}|\theta\rangle\langle\theta| +\frac {1} {8}(U \otimes
I)|\theta\rangle\langle\theta|(U^+ \otimes I)\nonumber\\
 +\frac {1} {8}(U \cdot U \otimes
I)|\theta\rangle\langle\theta|(U^+\cdot U^+ \otimes I)\nonumber\\
+\frac {1} {8}(U \cdot U \cdot U\otimes
I)|\theta\rangle\langle\theta|(U^+\cdot U^+ \cdot U^+\otimes
I)\nonumber\\ +\frac {1} {8}(V \otimes
I)|\theta\rangle\langle\theta|(V^+ \otimes I)\nonumber\\+\frac {1}
{8}(U \cdot V \otimes
I)|\theta\rangle\langle\theta|(V^+ \cdot U^+ \otimes I)\nonumber\\
+\frac {1} {8}(U \cdot U \cdot V \otimes
I)|\theta\rangle\langle\theta|(V^+\cdot U^+ \cdot U^+ \otimes I)\nonumber\\
+\frac {1} {8}(U \cdot U \cdot U \cdot V\otimes
I)|\theta\rangle\langle\theta|(V^+\cdot U^+ \cdot U^+ \cdot U^+
\otimes I).\nonumber
\end{eqnarray}
 With the matrix form, the mixed state can be
represented as
\begin{widetext}
\begin{eqnarray}
w=\frac {1} {4}\nonumber\\
\left(%
\begin{array}{cccccc}
  a^2+e^2+i^2+m^2 & ab+ef+ij+mn & ac+eg+ik+mr & ad+eh+iq+ms & (a+i)(e+m) & af+ej+bm+in \\
  ab+ef+ij+mn & b^2+f^2+j^2+n^2 & bc+fg+jk+nr & bd+fh+jq+ns & be+fi+jm+an & (b+j)(f+n) \\
  ac+eg+ik+mr & bc+fg+jk+nr & c^2+g^2+k^2+r^2 & cd+gh+kq+rs & ce+gi+km+ar & cf+gj+kn+br \\
  ad+eh+iq+ms & bd+fh+jq+ns & cd+gh+kq+rs & d^2+h^2+q^2+s^2 & de+hi+mq+as & df+hj+nq+bs \\
  (a+i)(e+m) & be+fi+jm+an & ce+gi+km+ar & de+hi+mq+as & a^2+e^2+i^2+m^2 & ab+ef+ij+mn \\
  af+ej+bm+in & (b+j)(f+n) & cf+gj+kn+br & df+hj+nq+bs & ab+ef+ij+mn & b^2+f^2+j^2+n^2 \\
  ag+ek+cm+ir & bg+fk+cn+jr & (c+k)(g+r) & dg+hk+qr+cs & ac+eg+ik+mr & bc+fg+jk+nr \\
  ah+dm+eq+is & bh+dn+fq+js & ch+gq+dr+ks & (d+q)(h+s) & ad+eh+iq+ms & bd+fh+jq+ns \\
  2(ai+em) & bi+aj+fm+en & ci+ak+gm+er & di+hm+aq+es & (a+i)(e+m) & be+fi+jm+an \\
  bi+aj+fm+en & 2(bj+fn) & cj+bk+gn+fr & dj+hn+bq+fs & af+ej+bm+in & (b+j)(f+n) \\
  ci+ak+gm+er & cj+bk+gn+fr & 2(ck+gr) & dk+cq+hr+gs & ag+ek+cm+ir & bg+fk+cn+jr \\
  di+hm+aq+es & dj+hn+bq+fs & dk+cq+hr+gs & 2(dq+hs) & ah+dm+eq+is & bh+dn+fq+js \\
  (a+i)(e+m) & af+ej+bm+in & ag+ek+cm+ir & ah+dm+eq+is & 2(ai+em) & bi+aj+fm+en \\
  be+fi+jm+an & (b+j)(f+n) & bg+fk+cn+jr & bh+dn+fq+js & bi+aj+fm+en & 2(bj+fn) \\
  ce+gi+km+ar & cf+gj+kn+br & (c+k)(g+r) & ch+gq+dr+ks & ci+ak+gm+er & cj+bk+gn+fr \\
  de+hi+mq+as & df+hj+nq+bs & dg+hk+qr+cs & (d+q)(h+s) & di+hm+aq+es & dj+hn+bq+fs \\
\end{array}%
\right. \nonumber
\end{eqnarray}
\begin{eqnarray}
\begin{array}{cccccc}
  ag+ek+cm+ir & ah+dm+eq+is & 2(ai+em) & bi+aj+fm+en & ci+ak+gm+er & di+hm+aq+es \\
  bg+fk+cn+jr & bh+dn+fq+js & bi+aj+fm+en & 2(bj+fn) & cj+bk+gn+fr & dj+hn+bq+fs \\
  (c+k)(g+r) & ch+gq+dr+ks & ci+ak+gm+er & cj+bk+gn+fr & 2(ck+gr) & dk+cq+hr+gs \\
  dg+hk+qr+cs & (d+q)(h+s) & di+hm+aq+es & dj+hn+bq+fs & dk+cq+hr+gs & 2(dq+hs) \\
  ac+eg+ik+mr & ad+eh+iq+ms & (a+i)(e+m) & af+ej+bm+in & ag+ek+cm+ir & ah+dm+eq+is \\
  bc+fg+jk+nr & bd+fh+jq+ns & be+fi+jm+an & (b+j)(f+n) & bg+fk+cn+jr & bh+dn+fq+js \\
  c^2+g^2+k^2+r^2 & cd+gh+kq+rs & ce+gi+km+ar & cf+gj+kn+br & (c+k)(g+r) & ch+gq+dr+ks \\
  cd+gh+kq+rs & d^2+h^2+q^2+s^2 & de+hi+mq+as & df+hj+nq+bs & dg+hk+qr+cs & (d+q)(h+s) \\
  ce+gi+km+ar & de+hi+mq+as & a^2+e^2+i^2+m^2 & ab+ef+ij+mn & ac+eg+ik+mr & ad+eh+iq+ms \\
  cf+gj+kn+br & df+hj+nq+bs & ab+ef+ij+mn & b^2+f^2+j^2+n^2 & bc+fg+jk+nr & bd+fh+jq+ns \\
  (c+k)(g+r) & dg+hk+qr+cs & ac+eg+ik+mr & bc+fg+jk+nr & c^2+g^2+k^2+r^2 & cd+gh+kq+rs \\
  ch+gq+dr+ks & (d+q)(h+s) & ad+eh+iq+ms & bd+fh+jq+ns & cd+gh+kq+rs & d^2+h^2+q^2+s^2 \\
  ci+ak+gm+er & di+hm+aq+es & (a+i)(e+m) & be+fi+jm+an & ce+gi+km+ar & de+hi+mq+as \\
  cj+bk+gn+fr & dj+hn+bq+fs & af+ej+bm+in & (b+j)(f+n) & cf+gj+kn+br & df+hj+nq+bs \\
  2(ck+gr) & dk+cq+hr+gs & ag+ek+cm+ir & bg+fk+cn+jr & (c+k)(g+r) & dg+hk+qr+cs \\
  dk+cq+hr+gs & 2(dq+hs) & ah+dm+eq+is & bh+dn+fq+js & ch+gq+dr+ks & (d+q)(h+s)\\
\end{array}%
\nonumber\\
\left.%
\begin{array}{cccccccccccccccc}
  (a+i)(e+m) & be+fi+jm+an & ce+gi+km+ar & de+hi+mq+as \\
  af+ej+bm+in & (b+j)(f+n) & cf+gj+kn+br & df+hj+nq+bs \\
  ag+ek+cm+ir & bg+fk+cn+jr & (c+k)(g+r) & dg+hk+qr+cs \\
  ah+dm+eq+is & bh+dn+fq+js & ch+gq+dr+ks & (d+q)(h+s) \\
  2(ai+em) & bi+aj+fm+en & ci+ak+gm+er & di+hm+aq+es \\
  bi+aj+fm+en & 2(bj+fn) & cj+bk+gn+fr & dj+hn+bq+fs \\
  ci+ak+gm+er & cj+bk+gn+fr & 2(ck+gr) & dk+cq+hr+gs \\
  di+hm+aq+es & dj+hn+bq+fs & dk+cq+hr+gs & 2(dq+hs) \\
  (a+i)(e+m) & af+ej+bm+in & ag+ek+cm+ir & ah+dm+eq+is \\
  be+fi+jm+an & (b+j)(f+n) & bg+fk+cn+jr & bh+dn+fq+js \\
  ce+gi+km+ar & cf+gj+kn+br & (c+k)(g+r) & ch+gq+dr+ks \\
  de+hi+mq+as & df+hj+nq+bs & dg+hk+qr+cs & (d+q)(h+s) \\
  a^2+e^2+i^2+m^2 & ab+ef+ij+mn & ac+eg+ik+mr & ad+eh+iq+ms \\
  ab+ef+ij+mn & b^2+f^2+j^2+n^2 & bc+fg+jk+nr & bd+fh+jq+ns \\
  ac+eg+ik+mr & bc+fg+jk+nr & c^2+g^2+k^2+r^2 & cd+gh+kq+rs \\
  ad+eh+iq+ms & bd+fh+jq+ns & cd+gh+kq+rs & d^2+h^2+q^2+s^2 \\
\end{array}%
\right).
\end{eqnarray}
\end{widetext}
The mutual information between Bob and Alice that can be extracted
from this state is given by the \emph{von-Neumann} entropy,
\emph{I}(Alice,Bob)$\leq S(w)=Tr\{wlog_2w\}$. In order to calculate
the \emph{von-Neumann} entropy, we need the eigenvalues $\lambda$ of
$w$, which are the roots of the characteristic polynomial det($w$).
Equivalently, we compute the roots of the characteristic polynomial
det($XwX^+$), yielding the 16 eigenvalues
\begin{eqnarray}
\lambda_{1,2}=\frac {1} {4}((a - i)^2 + (b - j)^2 + (c - k)^2 + (e -
m)^2 \nonumber\\
+ (f - n)^2 + (d - q)^2 + (g - r)^2 + (h - s)^2),\nonumber
\end{eqnarray}
\begin{eqnarray}
 \lambda_{3}=\frac {1} {4}((a - e + i - m)^2 + (b - f + j - n)^2\nonumber\\
  + (c - g + k -
r)^2 + (d - h + q - s)^2)),\nonumber\\
 \lambda_{4}=\frac {1} {4}((a + e + i + m)^2 + (b + f + j + n)^2\nonumber\\
  + ((c + g + k +
r)^2 +(d + h + q + s)^2))\nonumber,\\
 \lambda_{5-16}=0.\ \ \ \ \ \ \ \ \ \ \ \ \ \ \ \ \ \ \ \ \ \ \ \ \
 \ \ \ \ \ \ \ \ \ \ \ \ \ \ \ \ \ \
\end{eqnarray}
where
\begin{widetext}
\begin{eqnarray}
X=
\left(%
\begin{array}{cccccccccccccccc}
  \frac {1} {\sqrt {2}} & 0 & 0 & 0 & 0 & 0 & 0 & 0 & -\frac {1} {\sqrt {2}} & 0 & 0 & 0 & 0 & 0 & 0 & 0\\
  0 & \frac {1} {\sqrt {2}} & 0 & 0 & 0 & 0 & 0 & 0 & 0 & -\frac {1} {\sqrt {2}} & 0 & 0 & 0 & 0 & 0 & 0\\
  0 & 0 & \frac {1} {\sqrt {2}} & 0 & 0 & 0 & 0 & 0 & 0 & 0 & -\frac {1} {\sqrt {2}} & 0 & 0 & 0 & 0 & 0\\
  0 & 0 & 0 & \frac {1} {\sqrt {2}} & 0 & 0 & 0 & 0 & 0 & 0 & 0 & -\frac {1} {\sqrt {2}} & 0 & 0 & 0 & 0\\
  0 & 0 & 0 & 0 & \frac {1} {\sqrt {2}} & 0 & 0 & 0 & 0 & 0 & 0 & 0 & -\frac {1} {\sqrt {2}} & 0 & 0 & 0\\
  0 & 0 & 0 & 0 & 0 &\frac {1} {\sqrt {2}} & 0 & 0 & 0 & 0 & 0 & 0 & 0 & -\frac {1} {\sqrt {2}} & 0 & 0\\
  0 & 0 & 0 & 0 & 0 & 0 &\frac {1} {\sqrt {2}} & 0 & 0 & 0 & 0 & 0 & 0 & 0 & -\frac {1} {\sqrt {2}} & 0\\
  0 & 0 & 0 & 0 & 0 & 0 & 0 &\frac {1} {\sqrt {2}} & 0 & 0 & 0 & 0 & 0 & 0 & 0 & -\frac {1} {\sqrt {2}} \\
  \frac {1} {2} & 0 & 0 & 0 &-\frac {1} {2} & 0 & 0 & 0 & \frac {1} {2} & 0 & 0 & 0 &-\frac {1} {2} & 0 & 0 & 0 \\
  0 &\frac {1} {2} & 0 & 0 & 0 &-\frac {1} {2} & 0 & 0 & 0 & \frac {1} {2} & 0 & 0 & 0 &-\frac {1} {2}& 0 & 0\\
  0 & 0 &\frac {1} {2} & 0 & 0 & 0 &-\frac {1} {2} & 0 & 0 & 0 & \frac {1} {2} & 0 & 0 & 0 &-\frac {1} {2}& 0\\
  0 & 0 & 0 &\frac {1} {2} & 0 & 0 & 0 &-\frac {1} {2} & 0 & 0 & 0 & \frac {1} {2} & 0 & 0 & 0 &-\frac {1} {2}\\
  \frac {1} {2} & 0 & 0 & 0 &\frac {1} {2} & 0 & 0 & 0 & \frac {1} {2} & 0 & 0 & 0 &\frac {1} {2}& 0 & 0 & 0\\
  0 &\frac {1} {2} & 0 & 0 & 0 &\frac {1} {2} & 0 & 0 & 0 & \frac {1} {2} & 0 & 0 & 0 &\frac {1} {2}& 0 & 0\\
  0 & 0 &\frac {1} {2} & 0 & 0 & 0 &\frac {1} {2} & 0 & 0 & 0 & \frac {1} {2} & 0 & 0 & 0 &\frac {1} {2}& 0\\
  0 & 0 & 0 &\frac {1} {2} & 0 & 0 & 0 &\frac {1} {2} & 0 & 0 & 0 & \frac {1} {2} & 0 & 0 & 0 &\frac {1} {2}\\
\end{array}%
\right).
\end{eqnarray}
\end{widetext}
So we have
\begin{eqnarray}
I(Alice,Bob)\leq \nonumber\\
-\lambda_1log_2\lambda_1-\lambda_2log_2\lambda_2-\lambda_3log_2\lambda_3-\lambda_4log_2\lambda_4.
\end{eqnarray}
To $\lambda_1=\lambda_2=\lambda_3=\lambda_4=\frac {1} {4}$,
\emph{I}(Alice,Bob) reaches the maximal value 2 bits. So Bob can
eavesdrop 2 bits of 3 bits operation information on one two-qubit

Especially, \emph{I}(Alice,Bob) reaches the maximal value when Bob
prepares the legal two-qubit, namely $a=1$, or $ e=1$, or $i=1$, or
$m=1$, or $-a=e=i=m=\frac {1} {2}$, or $a=-e=i=m=\frac {1} {2}$, or
$a=e=-i=m=\frac {1} {2}$, or $a=e=i=-m=\frac {1} {2}$. Bob can not
gain more information by sending a fake single than by sending a
legal single.

Bob's eavesdropping introduces error into quantum state. Bob can not
gain Alice's input from the two-qubit operated by both Alice and her
next participant. Bob has to measure the two-qubit to extract
Alice's input before one two-qubit is resent. To eavesdrop 2 bits
operation information, Bob gains the maximal mixed state $\frac {1}
{4}I_4$ through sending a legal two-qubit, or gains one maximal
mixed state equivalent to $\frac {1} {4}I_4$ through sending a fake
singal. After extracting 2 bits operation information from $\frac
{1} {4}I_4$, he has to introduce $\frac {3} {8}$ error into whether
one reconstruction two-qubit or the collapse two-qubit.

\section{TROJAN HORSE ATTACK CAN BE DETECTED}
A Trojan horse attack bases on the idea that we can precisely know
an unknown quantum state by measuring many copies of the state.
Let Bob prepare the multi-qubit $\sum_{i_1i_2\ldots
i_m}a_{i_1i_2\ldots i_m}|i_1i_2\ldots i_m\rangle_{1,2,\ldots m}$
($m\geq 2$) to replace one data qubit $|d\rangle$.

In the detection scheme of the Trojan horse arrack (Fig. 1), Alice
prepares an auxiliary qubit $|a\rangle=|0\rangle$ or
$|a\rangle=|1\rangle$.

After the operation $H|a\rangle$, the system state is
\begin{eqnarray}
|\eta_1^0\rangle=\frac {1} {\sqrt {2}}\sum_{i_1i_2\ldots
i_m}a_{i_1i_2\ldots i_m}(|0\rangle+|1\rangle)|i_1i_2\ldots
i_m\rangle_{1,2,\ldots m}\nonumber\\
(|a\rangle=|0\rangle)\nonumber\\
 or \  |\eta_1^1\rangle=\frac {1} {\sqrt {2}}\sum_{i_1i_2\ldots
i_m}a_{i_1i_2\ldots i_m}(|0\rangle-|1\rangle)|i_1i_2\ldots
i_m\rangle_{1,2,\ldots m}\nonumber\\
(|a\rangle=|1\rangle).\nonumber\\
\end{eqnarray}
Here we use superscripts 0 and 1 to denote the states corresponding
to $a=0$ and $a=1$, respectively. This notation also applies to the
following equations and we will, for simplicity, suppress the word
``or" later.

Instead of the operator $C_{ad}$, the operators $C_{a1}, C_{a2},
\ldots, C_{am}$ are performed. The system state is
\begin{flushleft}
\begin{eqnarray}
|\eta_2^0\rangle=\frac {1} {\sqrt {2}}\sum_{i_1i_2\ldots
i_m}a_{i_1i_2\ldots i_m}(|0\rangle|i_1i_2\ldots
i_m\rangle_{1,2,\ldots m}\nonumber\\
+|1\rangle|\overline{i_1i_2\ldots
i_m}\rangle_{1,2,\ldots m}), \nonumber\\
|\eta_2^1\rangle=\frac {1} {\sqrt {2}}\sum_{i_1i_2\ldots
i_m}a_{i_1i_2\ldots i_m}(|0\rangle|i_1i_2\ldots
i_m\rangle_{1,2,\ldots m}\nonumber\\
-|1\rangle|\overline{i_1i_2\ldots i_m}\rangle_{1,2,\ldots m}).
\end{eqnarray}
\end{flushleft}
Instead of the operator $T_{ad}$, the operators $T_{a1}, T_{a2},
\ldots, T_{am}$ are performed. The system state is
\begin{eqnarray}
|\eta_3^0\rangle=\frac {1} {\sqrt {2^{m+1}}}\sum_{i_1i_2\ldots
i_m}|0\rangle|i_1i_2\ldots i_m\rangle_{1,2,\ldots
m}\nonumber\\
\{\sum_{x_2\ldots x_m}[(-1)^{\tau(i_1x_2\ldots x_m\oplus
i_1i_2\ldots i_m)}\nonumber\\
+(-1)^{\tau(i_1x_2\ldots x_m \oplus
\overline{i_1i_2\ldots i_m})}]a_{i_1x_2\ldots x_m}\}\nonumber\\
+\frac {1} {\sqrt {2^{m+1}}}\sum_{i_1i_2\ldots
i_m}|1\rangle|i_1i_2\ldots i_m\rangle_{1,2,\ldots
m}\nonumber\\
\{\sum_{x_2\ldots x_m}[(-1)^{\tau(i_1x_2\ldots x_m1\oplus
i_1i_2\ldots i_m0)}\nonumber\\
+(-1)^{\tau(i_1x_2\ldots x_m1 \oplus
\overline{i_1i_2\ldots i_m}0)}]a_{i_1x_2\ldots x_m}\}\nonumber\\
\nonumber\\
|\eta_3^1\rangle=\frac {1} {\sqrt {2^{m+1}}}\sum_{i_1i_2\ldots
i_m}|0\rangle|i_1i_2\ldots i_m\rangle_{1,2,\ldots
m}\nonumber\\
\{\sum_{x_2\ldots x_m}[(-1)^{\tau(i_1x_2\ldots x_m\oplus
i_1i_2\ldots i_m)}\nonumber\\
+(-1)^{\tau(i_1x_2\ldots x_m \oplus
\overline{i_1i_2\ldots i_m})+1}]a_{i_1x_2\ldots x_m}\}\nonumber\\
+\frac {1} {\sqrt {2^{m+1}}}\sum_{i_1i_2\ldots
i_m}|1\rangle|i_1i_2\ldots i_m\rangle_{1,2,\ldots
m}\nonumber\\
\{\sum_{x_2\ldots x_m}[(-1)^{\tau(i_1x_2\ldots x_m1\oplus
i_1i_2\ldots i_m0)}\nonumber\\
+(-1)^{\tau(i_1x_2\ldots x_m1 \oplus \overline{i_1i_2\ldots
i_m}0)+1}]a_{i_1x_2\ldots x_m}\},\nonumber
\end{eqnarray}
where $\overline{0}=1$, $\overline{1}=0$, and $\tau(x_1x_2\ldots
x_n)$ represents the number of $x_kx_{k+1}=11$ ($k=1,2,\ldots n-1$),
for example, $\tau(1100111)=3, \tau(1011011)=2$.

The detection scheme can detect a multi-qubit instead of a
single-qubit with probability $\frac {1} {2}$. When $m \geq 4$, the
probability amplitude of $|0\rangle|i_1\ldots
i_{m-3}i_{m-2}i_{m-1}i_m\rangle$ in $|\eta_3^0\rangle$  and that of
$|0\rangle|i_1\ldots
i_{m-3}\overline{i_{m-2}}i_{m-1}\overline{i_m}\rangle$ in
$|\eta_3^1\rangle$ are same or opposite, so the probability of
$|a\rangle=|0\rangle$ in the $|\eta_3^0\rangle$ equals to the
probability of $|a\rangle=|0\rangle$ in the $|\eta_3^1\rangle$.
Additionally verifying the cases of $m=2,3$, we can conclude that
when measuring the auxiliary qubit, if we gain  $|1\rangle$ in the
$|\eta_3^0\rangle$ with probability $\alpha$, we must gain
$|0\rangle$ in the $|\eta_3^1\rangle$ with probability $1-\alpha$.
So the auxiliary qubit flips with probability $\frac {1} {2}$.

Since the detection is a linear operation applied to quantum state,
it will work not only with pure states, but also with mixed states.
For example, Bob sends the legal state $|\overline{00}\rangle$ or
many copies of $|\overline{00}\rangle$ to Alice, the mixed state
inputs the detection. With two auxiliary qubits, Alice can detect
the case of many copies of $|\overline{00}\rangle$ with probability
$\frac {3} {4}$.

\section{CONCLUSION}
In this paper, we have presented a threshold quantum protocol based
on two-qubit operation. The number of qubits of the generated
quantum state by the threshold protocol equals to that of the
quantum state generated by the original (nonthreshold) protocol.
Fake signal attack  strategy and Trojan horse attack strategy of the
dishonest participant are investigated. The proposed protocol is
shown to resist these attacks.

The proposed two-qubit operation based on Grover's algorithm can
also be included in QSDC protocols and MQSS protocols. The proposed
detection scheme of Trojan horse attack can be included in the other
quantum cryptography protocols.

This work is supported by the National Natural Science Foundation of
China, Grants No. 60373059; the National Laboratory for Modern
Communications Science Foundation of China; the National Research
Foundation for the Doctoral Program of Higher Education of China,
Grants No. 20040013007; the Major Research plan of the National
Natural Science Foundation of China (90604023); and the ISN Open
Foundation.

\end{document}